# EVOLUTION OF EXCHANGE RATE REGIME:
# IMPACT ON MACRO ECONOMY OF BANGLADESH.

by

LIZA FAHMIDA

A thesis submitted in partial fulfillment of the requirements for the
degree of Professional Masters In Banking and Finance

Examination Committee:   Dr. Sundar Venkatesh (Chairperson)
                         Dr. Juthathip Jongwanich (Member).
                         Dr. Yousre Badir (Member)

Nationality:    Bangladeshi
Previous Degree:   Master in Finance and Banking
                   University of Dhaka
                   Bangladesh

Scholarship Donor:   Bangladesh Bank

Asian Institute of Technology
School of Management
Thailand
May 2012

# ACKNOWLEDGEMENT

The dissertation paper entitled *"Evolution Of Exchange Rate Regime: Impact On Macro Economy Of Bangladesh"* has been prepared for the partial fulfillment of Professional masters in Banking and Finance (PMBF) program conducted by School of Management, AIT, Thailand. I would like to offer my wholehearted gratitude and respect to a good number of people who offered encouragement, data and information, inspiration and assistance during the course of constructing this dissertation paper.

It would be difficult to prepare the paper and to present it in a lucid manner within stipulated time without the help of my guide teacher Dr. Sundar Venkatesh, Adjunct Faculty, School of Management, Asian Institute of Technology, Thailand. His utmost care, constant support and meticulous supervision guided me through the process.

I am indebted to Begum Sultana Razia, General Manager, Monetary Policy Department, Bangladesh Bank, whose sincere co-operation and valuable advice help me to prepare this. I would like to thank all of my colleagues and other employees of Foreign Exchange Policy Department, Forex Reserve and Treasury Management Department and Monetary Policy Department, Bangladesh Bank for their assistance in this regard.

I would like to thank all of fellow coursemates who contributed through their comments and suggestions to prepare the report in a comprehensive manner.

## ABSTRACT

Bangladesh had two different exchange rate regimes- a fixed exchange rate system from January 1972-May 2003 and a floating exchange rate regime since June 2003. After adopting the floating exchange rate regime Bangladesh experienced positive impacts on macro economic development. The variables of the macroeconomic factors have been considered as foreign reserve, workers' remittances and export proceeds. But the ongoing challenges for the country are the depreciating trend in local currency in a highly inflationary economy. The objective of the paper is to evaluate the macroeconomic performance over the regimes and to analyze present currency situation of Bangladesh.

**Table of Content**

# INTRODUCTION

At the preamble of the Bangladesh Bank order, 1972, it is stated that "Whereas it is necessary to establish a central bank in Bangladesh to manage the monetary and credit system of Bangladesh with a view to stabilizing domestic monetary value and maintaining a competitive external par value of the Bangladesh Taka toward fostering growth and development of country's productive resources in the national interest." To maintain a competitive external par value of the Bangladesh Taka, as per Foreign Exchange Regulation Act, 1947, Bangladesh Bank as a central bank of the country, regulates the foreign exchange on behalf of the government

## 1.1 Rationale:

Exchange rate indicates the global position of economy of the country. The country's economic development is closely related with it's foreign exchange system. Foreign exchange rate is a vital component for the country's economic activities too. Bangladesh has been experienced the floating exchange rate regime since May 2003 and passed a number of Pons and cons in the overall economy. Due to the utmost importance of the exchange rate in the economy, the study has been conducted in this area.

## 1.2 Scope:

This paper covers the comparison of the fixed and floating exchange regime of Bangladesh. The key factor of this paper is the evaluation of the impact of exchange rate on the fundamental macroeconomic indicators of the economy. Three main fundamental factors have been identified to measure their impact with exchange rate. Those are export, workers remittance and foreign exchange reserve. The empirical data of 2000 to 2012 has been used to justify the whole thing. To get an idea about the Bangladesh's position, some neighboring country's experience also been compared here.

## 1.3 Objectives of the study

2. To evaluate the exchange rate regimes in Bangladesh economy: Fixed and Floating
3. To evaluate the floating arrangement's performance in three macro economic variables: Export, Workers Remittance and Foreign Reserve.
4. To understand the reason for currency depreciation

## 1.4 Methodology

To obtain the objectives of the study, secondary data have collected. The sources of data are Bangladesh Bank, Board of Investment, Export Promotion Bureau, Websites of IMF, ADB, OANDA and other related links. Some regression analysis has been calculated to understand the exchange rate's significance on the economy.

## 1.5. Limitations

There were some limitations to conduct the study. Being, sensitive, new one, problems were faced to relate with various components and linking with them. And for its very nature, primary data was not available. Since the exchange rate regime is a vast area, it was also challenging to prepare this report within a limited time.

## 1.6. Organization

There are six chapters in this report. Chapter I is the introduction that divided into six sub sections. Literature Review is in Chapter 2. Exchange Rate Evaluation in Bangladesh has been covered in Chapter 3. Chapter 4 covers

Performance of Floating Exchange Rate System On Macro Economy of Bangladesh. Present exchange rate situation is describing in Chapter 5. The last chapter covers recommendation and conclusion.

# Chapter-02

## LITERATURE REVIEW

The exchange rate is one of the most important policy variables, which determines the trade flows, capital flows & FDI, inflation, international reserve and remittance of an economy. Many economies, specially Asian countries encountered crisis in 1990s due to imprudent application and bad choice of this policy. However, there is no consensus in the theoretical or empirical literature about any unique effect of the exchange rate volatility on macroeconomic indicators.

Whatever the case may be, different countries adopt different exchange rate policies. Bangladesh, the focus of this paper, had a fixed exchange rate system in place since January, 3 1972. After more than 31 years, the Central Bank of Bangladesh (Bangladesh Bank) changed it into a floating exchange rate system in June 2003.Bangladesh has been pursuing a floating exchange rate system since then. Dr. Mirza Azizul Islam, the former advisor, Ministry of Finance of the Caretaker Government of Bangladesh, presented a paper in January 2003, right before the shift from fixed to floating regime, explaining the overall performance of the fixed regime and the probable implications of the floating regime on Bangladesh economy. He suggested that the experiences of other countries in the region show that floating regime generates greater volatility in exchange rates and this sort of uncertainty is likely to affect adversely the overall trade and investment climate which is already afflicted by many unfavorable elements in Bangladesh (See Islam, 2002).

Bangladesh followed a 'fixed exchange rate' system until 1979. Between 1979 and mid-2003, the country pursued a managed floating exchange rate regime. Continual devaluation of the domestic currency, in order to maintain a stable real exchange rate and avoid overvaluation of the domestic currency, was the hallmarks of this regime. Since the end of May 2003, Bangladesh has introduced a kind of 'clean floating' exchange rate policy by making it fully convertible on the current account, although capital account controls still remain. All the exchange rate policies Bangladesh has taken, mainly, to accelerate exports, reduce extra pressure of imports and thereby improve the balance of trade. The following studies validate the above statement.

Islam (2002) states that the monetary authority determines the exchange rate policy aiming to achieve two main objectives. First, the 'domestic target', which includes restraining inflation rate, credit growth in the public and private sector, and the growth of liquidity and broad money. Secondly, the 'external target', which includes promotion in international reserves level, reduce the current account gap, control trends of exchange rate changes in the local inter-bank foreign exchange market, and adjust the trends in the exchange rates of neighboring trade partners: India, Pakistan and Sri Lanka.

Hossain et al (2000) quoted from Rahman (1995) and Bayes et al (1995) that the main objectives of exchange rate changes of Bangladesh were to: (i) promote international competitiveness; (ii) encourage exports diversification; (iii) withdraw subsidies from exports sector; (iv) discourage imports growth; and (v) rearrange resources in import substitutes and export oriented sectors. Aziz (2003) paper states that the finance ministers of last few regimes in their statement stated the following reason of devaluation of currency in Bangladesh: (i) increase export, (ii) discourage import, (iii) protect local infant industries, (iv) encourage the expatriates to send money to home, and (v) improve international reserve situation.

According to the Financial Sector Review (2006) of the Bangladesh Bank, the key aims of exchange rate policy of Bangladesh are to: (i) maintain competitiveness of Bangladeshi products in the world markets, (ii) encourage remittances inflow from expatriate wage earners,(iii) maintain stable internal price, and (iv) maintain a viable external account position. Thus, all the studies and policy papers have directly or indirectly articulated the export-led-growth and imports contraction targets as the main objectives of the exchange rate policy of Bangladesh.

Prior to adopting floating exchange rate regime, Islam (2002) argued that the economic and institutional prerequisites of a floating exchange rate regime are not met in Bangladesh. Some recent studies have tried to explain the behavior of nominal exchange rates of Bangladesh after its transition to the floating rate regime. By doing a correlation analysis, Rahman and Barua (2006) explore the possible explanation of the exchange rate movement. They found that there is a strong correlation (-0.40) between depreciation and export-import gap as a share of reserves; L/C openings for imports also have a positive correlation (0.45) with volatility of the exchange rate, which implies that the higher the L/C openings the more volatile is the exchange rate. They conclude that high seasonal demand for foreign currency because of increased import bills, systematic withdrawal of excess liquidity by Bangladesh Bank, relatively faster expansion of credit and higher interest rates on various national savings instruments are the reasons behind the interest rate hike in the money market and depreciation of the nominal exchange rate.

Miles, W. (2006) discussed about the effect of exchange rate system (both fixed and floating) on the long term growth in the economy. The effect of fixed rate, pegged rate and floating rate has been discussed here briefly on the basis of several literature reviews and the finding is that fixed and intermediate regimes have a clear, significantly negative impact on growth (holds only for emerging markets not for industrial nations.) In this paper, the author has given an approach to determine if exchange rate regime itself truly extracts an independent effect on growth. Results here indicate that the effect of fixed exchange rates on growth in emerging markets is not direct, but rather contingent on the existence of macroeconomic imbalances and other distortions in place in the domestic economy. These results seem to conform more closely with exchange rate theory, which posits mostly positive, and few negative channels for pegged currencies to impact growth over the long run.

Priyo (2009) examines whether the floating exchange rate regime has any impact on the value of Bangladesh taka i,e does it make any lose on the value of the currency. He shows that regime change has no statistically significant impact on the value of Bangladesh currency once foreign exchange reserve is incorporated in the regression model.

Chowdhury (2006) et al,. made an attempt to analyze Bangladesh's transition to floating regime and its impact on macroeconomic variables. They find that output growth in Bangladesh performed well in the intermediate and floating exchange rate regimes. Inflation is lower in the intermediate regime despite higher money supply and exchange rate depreciation. They also find that currency depreciation boosted export growth in the floating regime. Chowdhury et al,. (2006) have analyzed the exchange rate pass through to domestic inflation in Bangladesh.

The experiences of some countries in the region which implemented major changes in their exchange rate regimes in recent years can provide useful lessons for Bangladesh. I have tried to focus the comparison among different economic indicators between Bangladesh and some of the South Asian Countries (Afghanistan, Bhutan, India, Maldives, Nepal, Pakistan, and Sri Lanka) all of which adopted independently floating exchange rate regimes.

As an important trade and financial policy measure Bangladesh has changed its exchange rate policy accepting the obligations of IMF article VIII on March 24, 1994 by making the local currency taka fully convertible for current account transaction. Subsequently, as a member of IMF, Bangladesh was under pressure to open its exchange rate market. Finally, on May 31, 2003 Bangladesh introduced floating exchange rate system in current account. The IMF approved a loan for Bangladesh under the Poverty Reduction and Growth Facility (PRGF) only after switching to a floating exchange rate system (BB Website 2012).

Nevertheless, Chowdhury et al,. (2006) viewed that the free floating exchange rate regime was deployed in order to prevent the overvaluation of home currency because such overvaluation would make the exports less competitive in the world market and the import substitutes harder to compete with import goods. Islam (2002) states that the key objective of the free floating exchange rate regime is to prevent the major misalignment of exchange rate, particularly, to prevent any unexpected appreciation of real exchange rate which can harm exports demand, to promote the present export level and reduce the current account deficit, restrain inflation, and enhance the remittances.

Bangladesh has been pursued an active exchange rate policy from the very inception of the country's independence in 1971, which is reflected in frequently announced nominal exchange rate changes of the Bangladesh Bank. Islam (2002) cited 89 adjustments in the exchange rate of taka in terms of dollar from 1983 onward of which 83 were devaluation. Aziz (2008) observed 41 times devaluation between 1991 and 2000. Chowdhury et al,. (2006) state that between 1972 and 2002 the 'Taka' was devalued about 130 times to reduce the balance of payment deficits of Bangladesh. Hence this paper rightly chooses the exchange rate as the key explanatory variable for economic consequences of Bangladesh.

After independence, Bangladesh fixed its Taka's value with British Pound Sterling on 3rd 1972. Since 1972 to 1990 the Taka was overvalued. So there was a huge deficit in Balance of payment that hurt the economy badly. From 1990 the gap began to narrow down. The exchange rate regime worked quite well in terms of balance of payment, inflation, export and remittance. So there were some debate about incipience of this floating and criticism also rose about competence and preparation of Bangladesh Bank. But Bangladesh Bank performed well in managing the 'new born'. But there is some volatility in the market in recent past. Bangladesh Bank, as a central bank of the country intervened prudently to curve the volatility and market became stable though Taka remains undervalued.

Muhammad et al,. (2002) analyzed another important issue of whether stock prices and exchange rates are related or not has received considerable attention after the East Asian crises. They said that during the crises the countries affected saw turmoil in both currency and stock markets. If stock prices and exchange rates are related and the causation runs from exchange rates to stock prices then crises in the stock markets can be prevented by controlling the exchange rates. Moreover, developing countries can exploit such a link to attract/stimulate foreign portfolio investment in their own countries. Similarly, if the causation runs from stock prices to exchange rates then authorities can focus on domestic economic policies to stabilize the stock market. If the two markets/prices are related then investors can use this information to predict the behavior of one market using the information on other market. They also claimed that most of the empirical literature that has examined the stock prices-exchange rate relationship has focused on examining this relationship for the developed countries with very little attention on the developing countries.

This paper will assess whether the exchange rate regime change indeed has created any significant impact on the economy of the nation as well as the comparative analysis with the neighboring countries situation.

There are some potentialities as well as difficulties in market based system, so in this study there are some recommendations for the regulator and for the market players. To manage the floating exchange rate, full automation as well as transparency is essential in banking sector. As the capital account of our economy is not convertible there is little scope of capital flight. If the inception of floating exchange rate is the beginning to liberalize capital account immediately that won't be a wise decision.

# Chapter-3

## Exchange Rate Evolution in Bangladesh

### Introduction:

The world economy experienced some sort of fixed and flexible exchange rate. Before 1875 there was Bimetalism of exchange rate and then the Gold Standard (1875-1914). During Interwar period (1914-1944) the classical Gold Standard broken down and in July 1944 representatives of 44 countries succeeded to establish the 'Bretton Woods' system. Again the oil shock in the early 1970s and the dampen of demand broken down this system and world economy shifted to flexible exchange rate.

Bangladesh has been experienced two major exchange rate regimes since the country's Independence from $16^{th}$ December1971. A Fixed Exchange Rate Regime from 1972 to 1979 and a Floating Exchange Rate Regime since May 2003. Among the time frame from 1971 to 2003, there were different exchange rate arrangements in terms of the currency mechanism, like: Pegged to Pound Sterling (£):1972-1979;Pegged to a basket of major trading partners' currencies(£ as the intervening currencies):1980-1982;Pegged to a basket of major trading partners' currencies(US$ as the intervening currencies):1983-1999;Adjusted Pegged System:2000-2003;Floatig Exchange Rate System: May 30, 2003-Present. All the policies of exchange rate system Bangladesh implemented, with the objectives of accelerating exports, reducing import pressure and improve the balance of trade.

After independence, Bangladesh fixed its Taka's value with British Pound Sterling on $3^{rd}$ 1972. Since 1972 to 1990 the Taka was overvalued. So there was a huge deficit in Balance of payment that hurt the economy badly. From 1990 the gap began to narrow down. The exchange rate regime worked quite well in terms of balance of payment, inflation, export and remittance.

## 3.1 Factors Affecting Exchange Rate:

Under Floating Exchange Rate Arrangements Exchange Rate is primarily determined by demand for foreign currency and Supply of foreign currency where demand and supply of foreign currency is also affected by some other sensitive factors.

Theoretically demand for foreign currency is determined by several factors like, import payments, service payments which includes income payments, debt service payments, foreign a(outward) and foreign Investment (outward).

The supply of foreign currency is composed of export Receipt, service receipts which includes income receipts, debt service receipts, foreign aid (inward) and foreign Investment (inward).

Besides those some other factors affect the exchange rate movements. The factors are a)Purchasing Power parity; b)Interest rate parity; c) Relative income differential; d)Government Control; e)Expectations etc.

## 3.2 Performance of previous and current FX-Regime in Bangladesh

To evaluate the performance of Bangladesh considering the two different exchange rate regimes, some data comparison have been made among three neighboring countries in South Asia. Those are India, Pakistan and Sri Lanka.

### 3.2.1 Growth rate of GDP

Comparing the percentage of growth rate with major neighboring Countries, the following table shows that Bangladesh was more or less in a similar situation before the adoption of floating exchange rate regime. Since 2003 with the new floating exchange rate system, there is also a positive trend of the GDP growth rate except 2009-2010. The global recession affect the overall growth of the country at that time.

Table No.-3.1: Growth rate of GDP (% per year)

| Country | | | | Y | E | A | R | | | | |
| | 2001 | 2002 | 2003 | 2004 | 2005 | 2006 | 2007 | 2008 | 2009 | 2010 | 2011 |
|---|---|---|---|---|---|---|---|---|---|---|---|
| Bangladesh | 5.3 | 4.4 | 5.3 | 5.7 | 6.0 | *6.6* | *6.4* | *6.2* | *5.7* | *5.8* | *6.3* |
| India | 5.8 | 4.0 | 8.2 | 7.4 | 7.6 | 9.7 | 9.2 | 6.7 | 8.0 | 8.6 | 8.2 |
| Pakistan | 1.8 | 3.1 | 5.1 | 5.5 | 5.8 | 5.8 | 6.8 | 3.7 | 1.2 | 4.1 | 2.5 |
| Sri Lanka | -1.5 | 4.0 | 5.9 | 5.0 | 5.5 | 7.7 | 6.8 | 6.0 | 3.5 | 7.6 | 8.0 |

*Source: Asian Development Outlook-2004 and 2011, ADB*

**3.2.2 Current Account Balance**

In comparison to other major South Asian countries, the table no.-2 shows that Bangladesh's achievement in terms of containing current account balance is better after the adoption of floating exchange rate regime (since 2003). It has done consistently better than some of the neighboring countries like Sri Lanka, Pakistan, and India in all recent years excepting 2005.

Table No.3.2 Current Account Balance as Percentages of GDP

| Country | | | | Y | E | A | R | | | | | |
| | 1999 | 2000 | 2001 | 2002 | 2003 | 2004 | 2005 | 2006 | 2007 | 2008 | 2009 | 2010 | 2011 |
|---|---|---|---|---|---|---|---|---|---|---|---|---|---|
| Bangladesh | -1.5 | -1.1 | -2.3 | 0.4 | 0.5 | 0.0 | -1.5 | 1.3 | 1.4 | 0.9 | 2.7 | 3.7 | 0.2 |
| India | -1.1 | -0.8 | 0.2 | 0.8 | 0.7 | 0.3 | 0.3 | -1.2 | -1.4 | -2.4 | -2.8 | -3.0 | -3.5 |
| Pakistan | -3.0 | -0.4 | 0.6 | 4.6 | 5.9 | 3.0 | 2.1 | -3.9 | -4.8 | -8.5 | -5.7 | -2.2 | -1.7 |
| Sri Lanka | -3.6 | -6.4 | -1.5 | -1.8 | -2.2 | -3.0 | -3.5 | -5.3 | -4.3 | -9.5 | -0.5 | -3.8 | -4.0 |

*Source: Asian Development Outlook-2011, ADB*

### 3.2.3  Inflation Situation

Exchange rate regime and inflation are relevant because a change in the exchange rate is almost certain to cause a change in the domestic price of tradable and indirectly the price of non-tradable also. The international competitiveness of the economy is badly eroded by inflation. It generally encourages capital flight, exacerbates income distribution, gives rise to inequities in income distribution and aggravates poverty. The relevant data are presented in the following table no.-3.

TABLE-3.3:INFLATION IN BANGLADESH AND SELECTED SOUTH ASIAN COUNTRIES

| Country | Y E A R | | | | | | | | | | | | |
|---|---|---|---|---|---|---|---|---|---|---|---|---|---|
| | 1999 | 2000 | 2001 | 2002 | 2003 | 2004 | 2005 | 2006 | 2007 | 2008 | 2009 | 2010 | 2011 |
| Bangladesh | 8.9 | 3.4 | 1.6 | 2.8 | 4.4 | 4.7 | 5.2 | 7.2 | 7.2 | 9.9 | 6.7 | 7.3 | 8.0 |
| India | 3.3 | 7.2 | 4.7 | 3.4 | 5.3 | 5.0 | 5.0 | 5.2 | 5.0 | 8.7 | 2.1 | 9.2 | 7.8 |
| Pakistan | 5.7 | 3.6 | 4.4 | 3.5 | 3.1 | 4.0 | 6.2 | 7.9 | 7.8 | 12.0 | 20.8 | 11.7 | 16.0 |
| Sri Lanka | 5.9 | 1.2 | 11.0 | 10.2 | 2.6 | - | - | 10.0 | 15.8 | 22.6 | 3.4 | 5.9 | 8.0 |

*Source: Asian Development Outlook-2011, ADB*
It is showing that there is an increasing trend of inflation since the adaptation of floating exchange rate regime. Data on inflation rates represent period averages. Except for India, which reports the wholesale price index, inflation rates presented are based on consumer price indexes. The higher inflationary situation take place due oil and food price hike in the international market as well as several natural disaster like, Sidor, flood etc.

### 3.3 Justification of Floating Exchange Rate:

From the above mentioned data analysis, it can be said the previous regime performed quite well in certain criteria. The major reasons behind the adoption of new exchange rate system is mainly the government's commitment to the liberalization of the country's economy and to take the appropriate steps to create suitable environment of the economy for entering into capital account convertibility regime. Rather than this, there was IMF's 'conditionalities' to enter into new floating exchange rate regime.

## 3.4 The Transition from Fixed to Float:

To meet up the economic demand and to fulfill the aid conditionality, on 29 May, 2003 Bangladesh Bank issued a circular stating- effective from 31$^{st}$ May, 2003, Bangladesh Bank floated its exchange rate and followed a fully market based exchange rate for Taka. Under this arrangement, exchange rate is determined on the basis of demand and supply of the respective currencies.

Immediately after the inception of floating exchange rate banks, economists, currency traders and businessmen have welcomed the deregulation of the exchange rate saying that the country's foreign trade and remittance would get a boast up due to it and it would make the currency market more efficient and effective.

Since the introduction there is no unusual raise of exchange rate till mid 2004. Most of the time Taka maintains appreciating position during this period and Bangladesh Bank show a tremendous performance managing the 'new born' exchange rate system. During mid 2004, Taka faced significant volatility against USD and it continued up to August 2004. After that period, the volatility of exchange rate of Taka against USD eased but resulting to appreciated USD till mid January 2005. In recent times, Taka has depreciated significantly against USD in the inter-bank market. This has happened as because of higher growth in lending for the private sectors following devastating flood situation. Consequent on such higher growth, the private sector required huge importation of capital machinery and intermediate goods. Besides, price hike in oil price and scrap vessel in the international market created a surge on import settlements.

Figur-3.4: Exchange Rate Movement immediately after inception of Floating

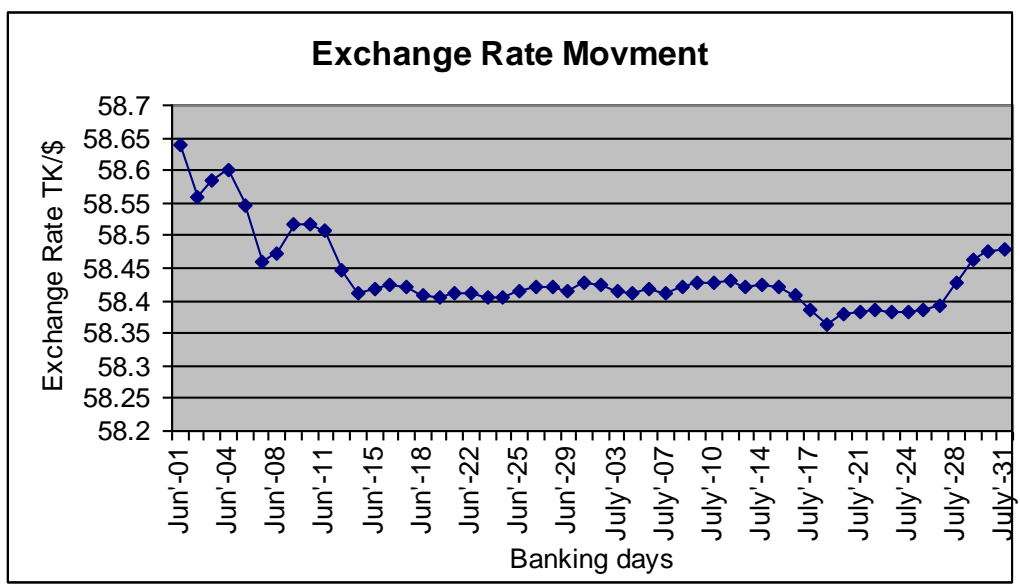

**Behavior of Real Effective Exchange Rate (REER) and Nominal Exchange Rate:**

REER is considered as a guiding FX-rate to the policy maker as well as the market participants and it also shows the international competitiveness of countries goods and services. The policy makers are always tried to keep the Nominal Exchange Rate near to the REER. Before 1990 Taka was overvalued (see table no-01,02 and 03.) that distorted our international competitiveness and that was the causes of prolonged Balance of Payment crisis.

Figure:3.5: Comparison of month end Real Effective Exchange Rate (REER) and Nominal Exchange Rate between 2003-2004.

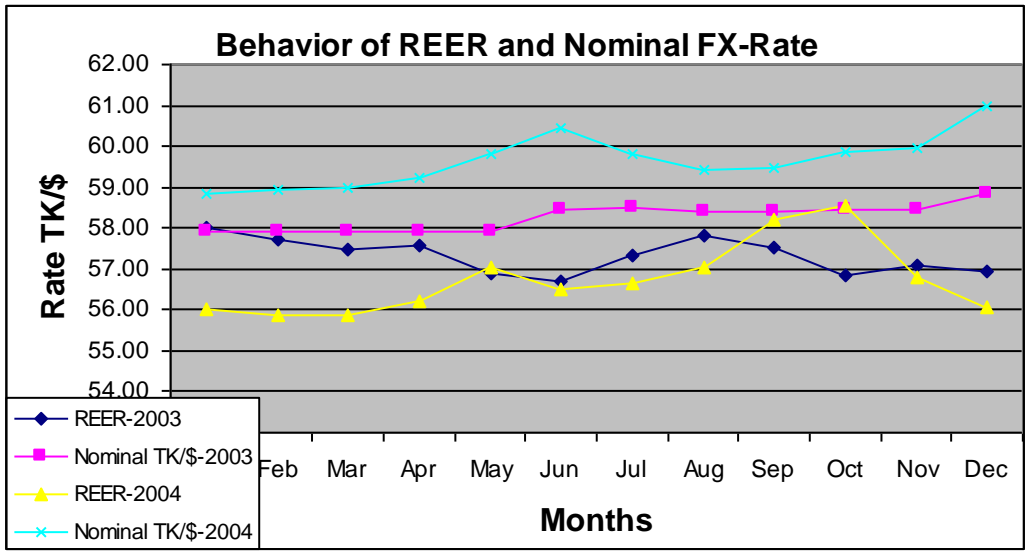

*Source: Bangladesh Bank Unpublished Data*

After 1990, Taka remained undervalued. The gap between Nominal Exchange Rate and REER were widened over the period.

Before inception of floating exchange rate Taka was almost running on a free float for the last 12-15 month as rates were decided according to demand-supply situation and liquidity in market. Bangladesh Bank was also not virtually selling any dollar that time. This means the market has already factored in floating exchange rate. Moreover, before inception Bangladesh Bank took some measures for strengthening regulation, such as, keeping close observation of Authorized Dealer's (AD's) daily activities; especially on 'Open Position', gave pressure for reconciliation of NOSTRO account balance, encourage every bank to set up dealing room and pressurized to make payment in due time etc. So we can obviously say that way to the floating exchange rate was prepared.

# Chapter - 4

## Performance of Floating Exchange Rate System
## On Macro Economy of Bangladesh

Introduction of floating exchange rate was debatable issue and also there were some criticisms about the competence of Bangladesh Bank's from some corner. But Bangladesh Bank performed a tremendous performance. There was no volatility; no speculation in price and market behaves rationally. If we consider the market statistics, we find that macro economic variables have positive performances over the period of time. Three major variables have been considered for evaluating the impact of exchange rate with them. The variables are Export, Workers Remittances and Foreign reserve.

### 4.1 The Export Situation:

The export trend from 1998 to 2011 shows an increasing trend. It is shown here that there is upward trend of export after 2003,i.e, after adopting the floating exchange rate regime, the export has a robust growth in the economy.

Figure:4.1: The Export Volume in US$

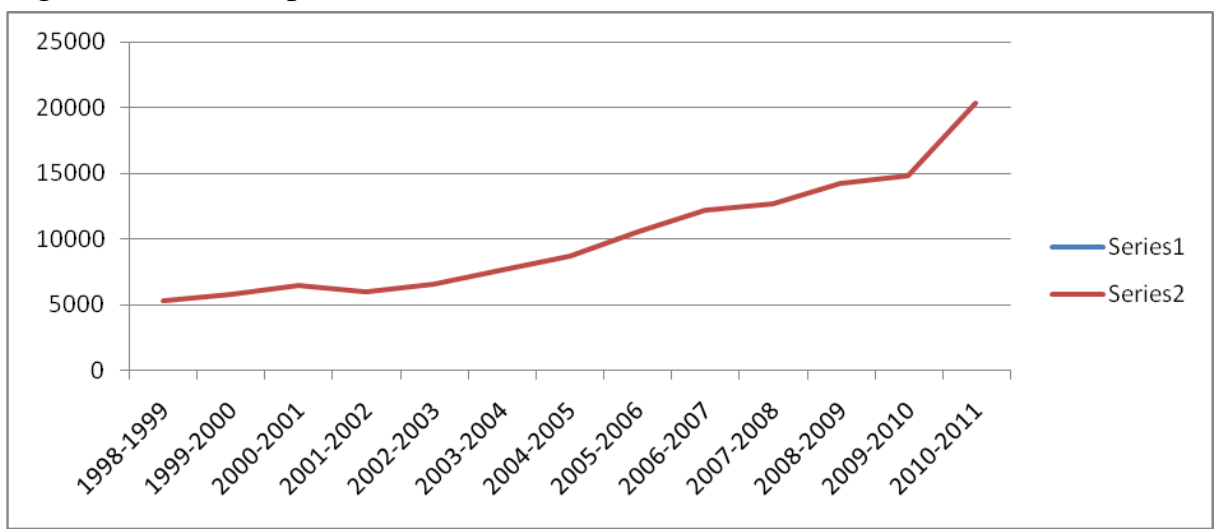

*Source: Bangladesh Bank Quarterly, January-March 2011*

During the global recession, the export trend of Bangladesh was not that much affected mostly for the RMG sector. In the FY 09 and FY 10, the export volume increased significantly.

Trade Deficit: Though the export volume has an increasing trend since the inception of floating exchange rate regime, there is huge amount of trade deficit with an increasing trend.

Figure:4.2-Trade Deficit of Bangladesh

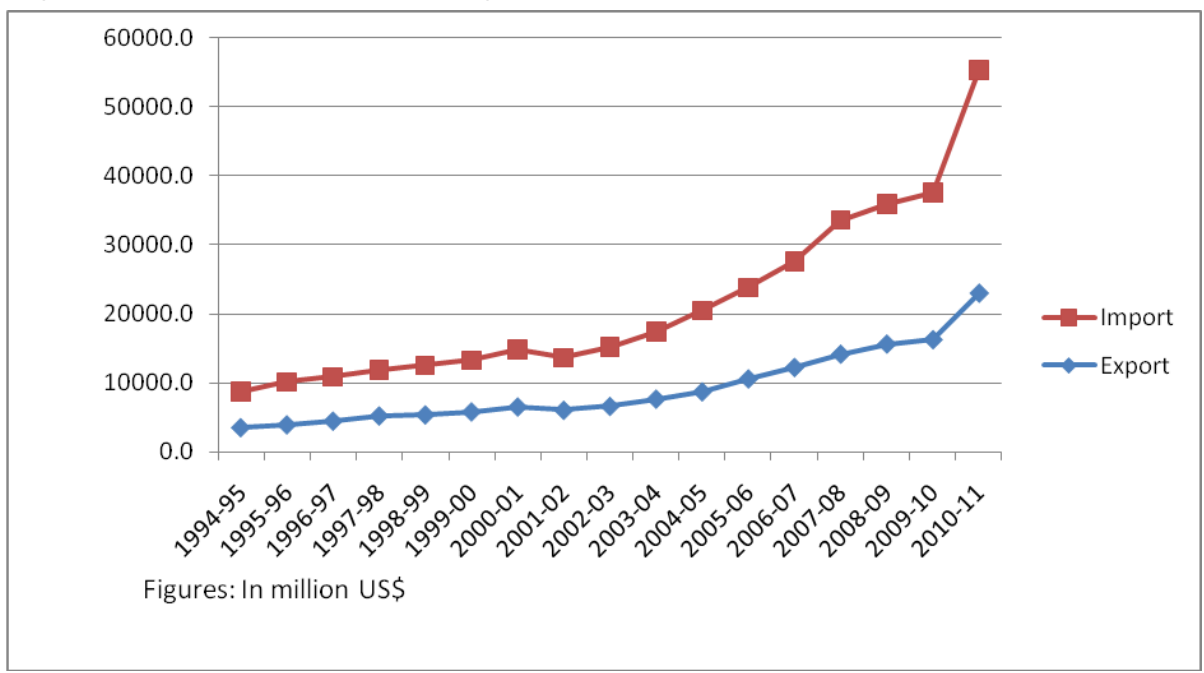

Source:Monthly Economic Trend-Bangladesh Bank , January 2012

The amount of import has been increased sharply after 2006-2007. The gap between export and import becomes huge during the last financial year 2010-2011.

## 4.2 Enhancement of Workers Remittance:

The Inward remittances from Bangladeshi nationals working abroad remained strong in FY10 even in the face of global economic slowdown and continued to play an important role in strengthening the current account. Receipts on this sector increased by 13.4 percent to USD 10987.40 million in FY10 from USD 9689.26 million in FY09. The underlying reason was that Bangladesh Bank has simplified the approval policy of drawing arrangements between foreign exchange houses and domestic banks. As a result, 40 banks have been allowed for establishing 885 drawing arrangements with 300 exchange houses all over the world for collecting remittances, (of which approximately 650 drawing arrangements with 250 exchange houses are operative now).\

Figure-4.3: The Workers Remittance

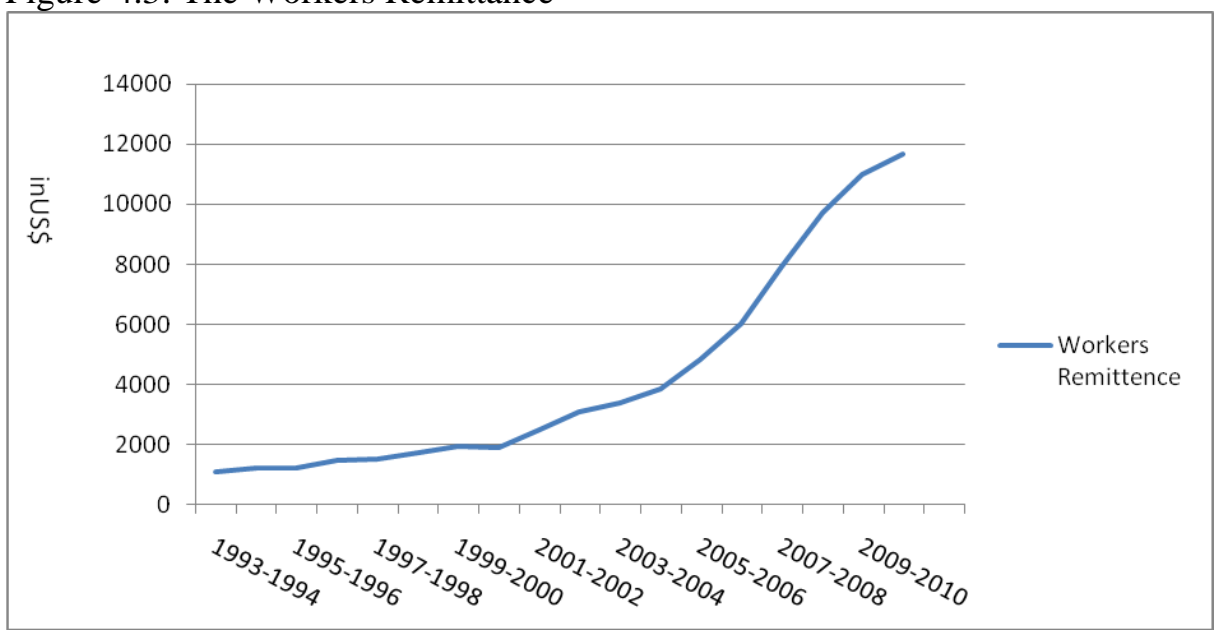

*Source: Bangladesh Bank Quarterly, January-March 2011*

Considering the growth rate of workers' remittances, it has been observed that the rate is quite higher after the free floating exchange rate regime that is 20.52 % (2003-2010) than that of fixed exchange rate regime of Bangladesh which is calculated as 11.89% (1993-2002). The increasing amount of workers remittance helps to balance the trade deficit in a prudent manner.

## 4.3 Reserve Position:

The amount of foreign exchange reserve has been increased significantly over the last couple of years. During the FY 2003 to FY 2005, the trend was quite steady and flat. But, it has an upward trend after FY 2007. The main sources of foreign reserve are workers remittance, foreign loans and grants and exports.

Figure-4.4: The Foreign Reserve Position

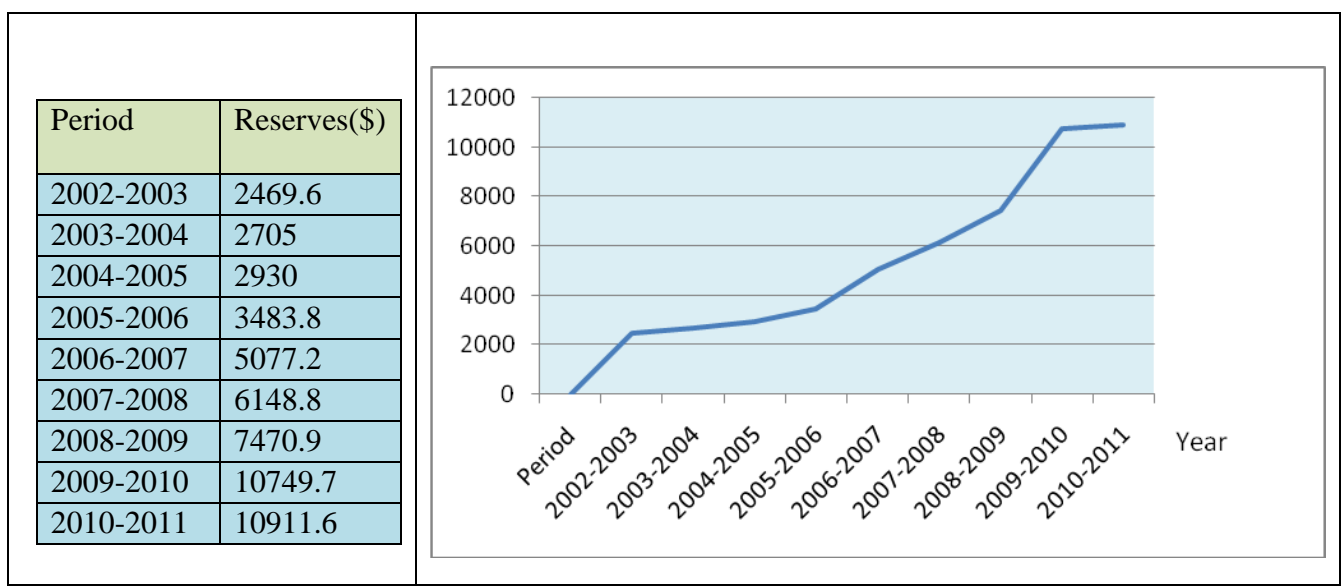

| Period | Reserves($) |
|--------|-------------|
| 2002-2003 | 2469.6 |
| 2003-2004 | 2705 |
| 2004-2005 | 2930 |
| 2005-2006 | 3483.8 |
| 2006-2007 | 5077.2 |
| 2007-2008 | 6148.8 |
| 2008-2009 | 7470.9 |
| 2009-2010 | 10749.7 |
| 2010-2011 | 10911.6 |

*Source: Bangladesh Bank Quarterly, January-March 2011*

After the inception of floating exchange rate regime, the foreign exchange reserve boosted up due to huge amount of workers remittance and increasing trend of export.

## 4.4 The Growth Rate of GDP:

The GDP growth rate reaches upto 6.7% during FY 11. From 1994 to 2010, the average growth rate of GDP was 5.47% reaching at high of 6.63% in June 2006. The record low rate was 4.08% during June 2004. For the last couple of years the growth rate was 5% above and Bangladesh is considered as a developing country.

Figure-4.5: Bangladesh GDP Growth Rate

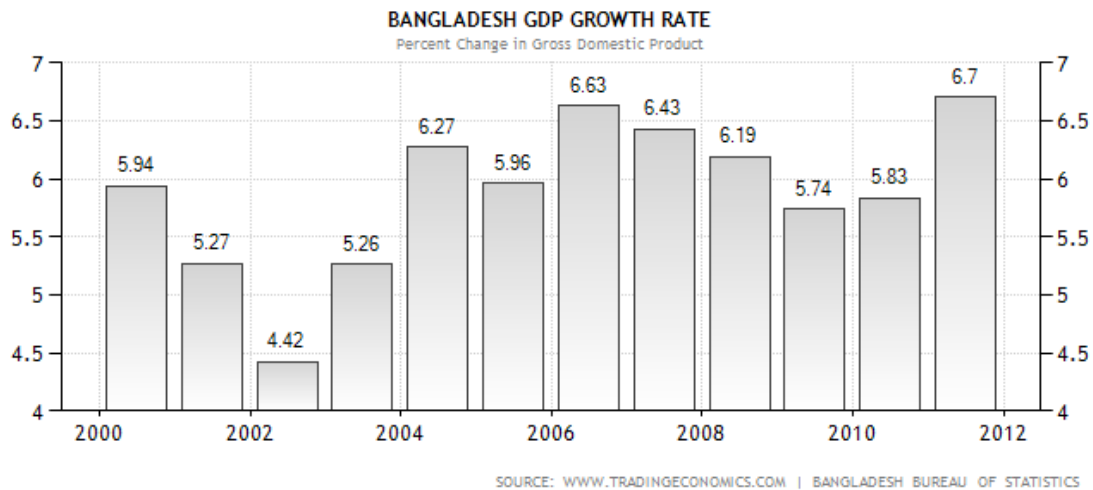

*Source: Bangladesh Bureau of Statistics,2011*

## 4.5 The relationship among the variables:

The relationship between the exchange rate and three variables i,e, export, remittance and reserve reflects quite positive in correlation. The correlation has been computed considering the data between 2002-2011. The correlation of remittance with exchange rate has been calculated as 0.859373146 which is very strong. For export, the correlation between exchange rate is 0.928315219 reflects very strong relationship. The correlation between reserve and exchange rate is calculated as 0.825444493which also very positive.

It is obvious from the calculation that the floating exchange rate regime has a positive impact on macro economic variables of the country.

<div align="center">

Chapter 5

**Preset Exchange Rate Situation**

</div>

In the recent Taka depreciates against US$ drastically: Over the period, July 2010-January 2012, from Tk 70 to Tk 86 -- a depressing 23% fall down.

Figure-5.1: Exchange Rate Movement

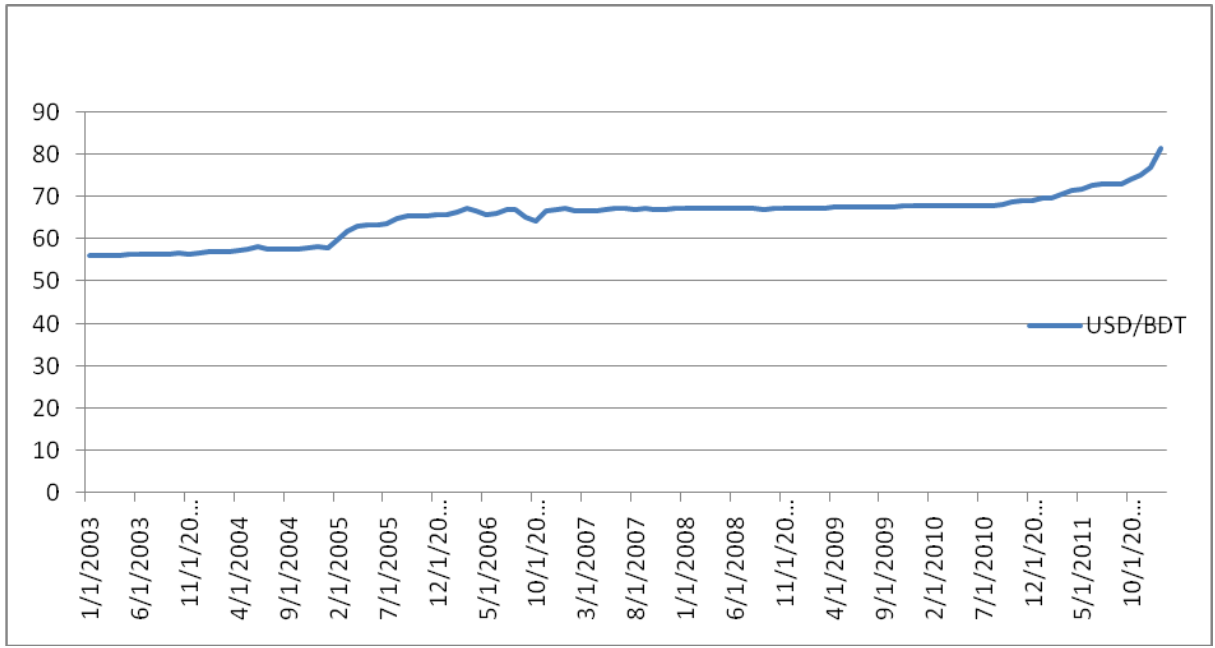

Source: Website OANDA

The sharp fall of taka against US dollar continues for last couple of months. US dollar has been weakening against many other strong currencies like Euro, SF, Yen and GBP but getting stronger against Bangladesh Taka. Exchange rate depreciation creates the economy in a challenging situation. Continuous depreciating tendency higher the inflation rate that ultimately increase the trade deficit.

After inception of the floating exchange rate system, the relationship between the exchange rate and the macroeconomic fundamentals showed that there is a positive relationship. But the ongoing depreciation of Bangladeshi taka is becoming the challenge for the central bank as well as the Government. The increasing trend of the trade deficit (figure: 7) also shows the possibility of the shortage of supply of the foreign currency i,e, US$ in the market.

The reasons behind the currency depreciation depends on economic fundamental, exchange rate regime and Trading rules. The economic fundamentals covers the necessity of foreign direct investment which ensure long term capital inflow in the country.

**5.1 Foreign Direct Investment:**

The growth rate of foreign direct investment is showing a declining trend. The data has been used from 1996 to 2011.

Figure-5.2: Growth Rate of Foreign Direct Investment

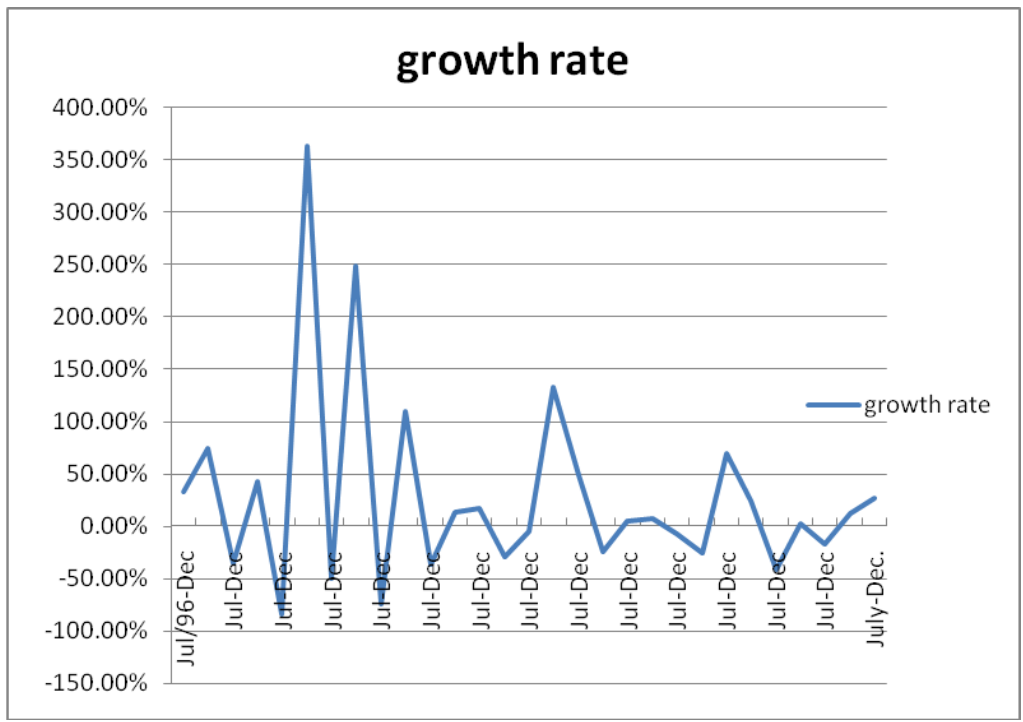

Source: Monthly Economic Trend-Bangladesh Bank, January 2012

In the recent past, the FDI growth rate is severely low. So, proper steps are supposed to be taken by the government authority .

Chapter-6

**Recommendations and Conclusion**

Considering the above mentioned discussions, some recommendations have been formulated. Those are discussing as follows:

**6.1 Enhancing Foreign Direct Investment:** Sufficient inflow of foreign direct investment could enhance the capital inflow in the country for long term. Bangladesh Government has specific policy for inviting FDI. Ensuring good governance, Infrastructural development, Utility, political stability will help to promote more investment from abroad in the country.

**6.2 Short Term Foreign Borrowings:** Borrowings from abroad is another option for supplying liquidity in the market. But the problem of short-term borrowing is that the country may fall into "Debt-trap" due to be unable to pay the money on time. Once the foreign currency injects in the market, it is difficult to recollect form the market as well. The Central Bank of Philippines in early 1990 can be remembered here. Due to short-term borrowing to meet the local market demand, got huge foreign liability. Continuous losses eroded it's capital base and made it bankrupt in 1993. It took 25 years to reestablish the new central bank in that country.

**6.3 Derivatives:** Introduction of various derivative products as options, currency swap, interest cap, interest swap, futures, forwards etc, as well as ensuring huge portfolio investment might bring the positive impact in the exchange rate market.

**6.4 Formation of Domestic FX Market:** A formal forex market forum should be created, with the participation of independent professional bodies and with representation from the dealers association, Bangladesh Bank and other relevant government officials. This forum would provide the logistic support and platform for the forex market. Ideally, this forum would develop a secured web-based market to which only the members or participating organizations would have access. From this website, all the logistic support required for completing the forex dealings among the members/participating organisations could be provided.

## CONCLUSION:

This study shows that floating exchange rate is affecting country's economy positively.  But in present system, for the ongoing depreciating trend of taka, the responsibility obviously goes to the regulator's shoulder. So the study recommended some sorts of measure for proper regulation. The regulation should be proactive rather than reactive. There are some potentialities in this new regime; to reap this potentiality, Government as well as regulators should be play role carefully.